\pdfoutput=1
\documentclass[12pt,a4paper]{article}

\usepackage{ifthen} 
\newboolean{pdflatex}
\setboolean{pdflatex}{true} 

\newboolean{articletitles}
\setboolean{articletitles}{true} 

\newboolean{uprightparticles}
\setboolean{uprightparticles}{false} 

\newboolean{inbibliography}
\setboolean{inbibliography}{false} 

\def\paperauthors{LHCb collaboration} 
\def\paperasciititle{} 
\def\papertitle{Search for the\\rare hadronic decay \BSpp} 
\def\paperkeywords{{High Energy Physics}, {LHCb}} 
\def\papercopyright{CERN on behalf of the LHCb collaboration}
\def\paperlicence{CC BY 4.0 licence}
\def\paperlicenceurl{https://creativecommons.org/licenses/by/4.0/}

\usepackage[top=1in, bottom=1.25in, left=1in, right=1in]{geometry}

\usepackage{microtype}
\usepackage{lineno}  
\usepackage{xspace} 
\usepackage{caption} 

\usepackage{graphicx}  
\usepackage{color}
\usepackage{colortbl}
\graphicspath{{./figs/}} 

\usepackage{amsmath} 
\usepackage{amssymb}
\usepackage{amsfonts}
\usepackage{upgreek} 

\newcommand*\patchAmsMathEnvironmentForLineno[1]{%
\expandafter\let\csname old#1\expandafter\endcsname\csname #1\endcsname
\expandafter\let\csname oldend#1\expandafter\endcsname\csname
end#1\endcsname
 \renewenvironment{#1}%
   {\linenomath\csname old#1\endcsname}%
   {\csname oldend#1\endcsname\endlinenomath}%
}
\newcommand*\patchBothAmsMathEnvironmentsForLineno[1]{%
  \patchAmsMathEnvironmentForLineno{#1}%
  \patchAmsMathEnvironmentForLineno{#1*}%
}
\AtBeginDocument{%
\patchBothAmsMathEnvironmentsForLineno{equation}%
\patchBothAmsMathEnvironmentsForLineno{align}%
\patchBothAmsMathEnvironmentsForLineno{flalign}%
\patchBothAmsMathEnvironmentsForLineno{alignat}%
\patchBothAmsMathEnvironmentsForLineno{gather}%
\patchBothAmsMathEnvironmentsForLineno{multline}%
\patchBothAmsMathEnvironmentsForLineno{eqnarray}%
}

\usepackage{hyperxmp}

\usepackage[pdftex,
            pdfauthor={\paperauthors},
            pdftitle={\paperasciititle},
            pdfkeywords={\paperkeywords},
            pdfcopyright={Copyright (C) \papercopyright},
            pdflicenseurl={\paperlicenceurl}]{hyperref}

\usepackage[colorinlistoftodos,textsize=scriptsize]{todonotes}

\usepackage[bottom,flushmargin,hang,multiple]{footmisc}

\usepackage[all]{hypcap} 

\usepackage{xspace} 
\usepackage{upgreek}

\def\lhcb   {\mbox{LHCb}\xspace}

\def\babar  {\mbox{BaBar}\xspace}
\def\belle  {\mbox{Belle}\xspace}

\def\cleo   {\mbox{CLEO}\xspace}

\def\aleph  {\mbox{ALEPH}\xspace}
\def\delphi {\mbox{DELPHI}\xspace}

\def\MagUp {\mbox{\em Mag\kern -0.05em Up}\xspace}

\def\lone   {L0\xspace}

\ifthenelse{\boolean{uprightparticles}}%
{

 \def\Ppi         {\ensuremath{\uppi}\xspace}

 \def\PDelta      {\ensuremath{\Delta}\xspace}                 
 \def\PXi         {\ensuremath{\Xi}\xspace}                 
 \def\PLambda     {\ensuremath{\Lambda}\xspace}                 
 \def\PSigma      {\ensuremath{\Sigma}\xspace}                 
 \def\POmega      {\ensuremath{\Omega}\xspace}                 
 \def\PUpsilon    {\ensuremath{\Upsilon}\xspace}
 \let\oldPi\Pi
 \def\PPi         {\ensuremath{\oldPi}\xspace}

 \def\PB      {\ensuremath{\mathrm{B}}\xspace}                 
                  
 \def\PD      {\ensuremath{\mathrm{D}}\xspace}

 \def\PK      {\ensuremath{\mathrm{K}}\xspace}

 \def\PW      {\ensuremath{\mathrm{W}}\xspace}

 \def\Pb      {\ensuremath{\mathrm{b}}\xspace}                 
 \def\Pc      {\ensuremath{\mathrm{c}}\xspace}

 \def\Ph      {\ensuremath{\mathrm{h}}\xspace}                 
 \def\Pi      {\ensuremath{\mathrm{i}}\xspace}

 \def\Pp      {\ensuremath{\mathrm{p}}\xspace}

 \def\Ps      {\ensuremath{\mathrm{s}}\xspace}                 
                  
 \def\Pu      {\ensuremath{\mathrm{u}}\xspace}

 \def\thebaroffset{0.0em}
}
{

 \def\Ppi         {\ensuremath{\pi}\xspace}

 \mathchardef\PDelta="7101
 \mathchardef\PXi="7104
 \mathchardef\PLambda="7103
 \mathchardef\PSigma="7106
 \mathchardef\POmega="710A
 \mathchardef\PUpsilon="7107
 \mathchardef\PPi="7105
                  
 \def\PB      {\ensuremath{B}\xspace}                 
                  
 \def\PD      {\ensuremath{D}\xspace}

 \def\PK      {\ensuremath{K}\xspace}

 \def\PW      {\ensuremath{W}\xspace}

 \def\Pb      {\ensuremath{b}\xspace}                 
 \def\Pc      {\ensuremath{c}\xspace}

 \def\Ph      {\ensuremath{h}\xspace}                 
 \def\Pi      {\ensuremath{i}\xspace}

 \def\Pp      {\ensuremath{p}\xspace}

 \def\Ps      {\ensuremath{s}\xspace}                 
                  
 \def\Pu      {\ensuremath{u}\xspace}

 \def\thebaroffset{0.18em}
}
\newcommand{\offsetoverline}[2][\thebaroffset]{\kern #1\overline{\kern -#1 #2}}%

\makeatletter
\ifcase \@ptsize \relax
  \newcommand{\miniscule}{\@setfontsize\miniscule{4}{5}}
\or
  \newcommand{\miniscule}{\@setfontsize\miniscule{5}{6}}
\or
  \newcommand{\miniscule}{\@setfontsize\miniscule{5}{6}}
\fi
\makeatother

\DeclareRobustCommand{\optbar}[1]{\shortstack{{\miniscule (\rule[.5ex]{1.25em}{.18mm})}
  \\ [-.7ex] $#1$}}

\def\W      {{\ensuremath{\PW}}\xspace}

\def\uquark    {{\ensuremath{\Pu}}\xspace}

\def\squark    {{\ensuremath{\Ps}}\xspace}

\def\cquark    {{\ensuremath{\Pc}}\xspace}

\def\bquark    {{\ensuremath{\Pb}}\xspace}

\def\pion   {{\ensuremath{\Ppi}}\xspace}

\def\pip    {{\ensuremath{\pion^+}}\xspace}
\def\pim    {{\ensuremath{\pion^-}}\xspace}

\def\kaon    {{\ensuremath{\PK}}\xspace}

\def\KorKbar {\kern \thebaroffset\optbar{\kern -\thebaroffset \PK}{}\xspace}

\def\Kp      {{\ensuremath{\kaon^+}}\xspace}
\def\Km      {{\ensuremath{\kaon^-}}\xspace}

\def\D       {{\ensuremath{\PD}}\xspace}

\def\DorDbar {\kern \thebaroffset\optbar{\kern -\thebaroffset \PD}\xspace}

\def\Dp      {{\ensuremath{\D^+}}\xspace}
\def\Dm      {{\ensuremath{\D^-}}\xspace}

\def\DpDm    {\ensuremath{\Dp {\kern -0.16em \Dm}}\xspace}

\def\B       {{\ensuremath{\PB}}\xspace}

\def\BorBbar {\kern \thebaroffset\optbar{\kern -\thebaroffset \PB}\xspace}

\def\Bd      {{\ensuremath{\B^0}}\xspace}

\def\BdorBdbar {\kern \thebaroffset\optbar{\kern -\thebaroffset \Bd}\xspace}
\def\Bu      {{\ensuremath{\B^+}}\xspace}

\def\Bp      {{\ensuremath{\Bu}}\xspace}

\def\Bs      {{\ensuremath{\B^0_\squark}}\xspace}

\def\BsorBsbar {\kern \thebaroffset\optbar{\kern -\thebaroffset \Bs}\xspace}

\def\Bds     {{\ensuremath{\B_{(\squark)}^0}}\xspace}

\def\Y#1S{\ensuremath{\PUpsilon{(#1S)}}\xspace}

\def\proton      {{\ensuremath{\Pp}}\xspace}
\def\antiproton  {{\ensuremath{\overline \proton}}\xspace}

\def\Lz          {{\ensuremath{\PLambda}}\xspace}
\def\Lbar        {{\ensuremath{\offsetoverline{\PLambda}}}\xspace}
\def\LorLbar     {\kern \thebaroffset\optbar{\kern -\thebaroffset \PLambda}\xspace}

\def\Lb           {{\ensuremath{\Lz^0_\bquark}}\xspace}

\def\BF         {{\ensuremath{\mathcal{B}}}\xspace}
\def\BR         {\BF}

\newcommand{\decay}[2]{\ensuremath{#1\!\to #2}\xspace} 

\def\to                 {\ensuremath{\rightarrow}\xspace}

\def\BTohh        {\decay{\B}{\Ph^+ \Ph'^-}}

\def\AT#1     {\ensuremath{A_{\mathrm{T}}^{#1}}\xspace}           

\def\C#1      {\ensuremath{\mathcal{C}_{#1}}\xspace}                       
\def\Cp#1     {\ensuremath{\mathcal{C}_{#1}^{'}}\xspace}                    
\def\Ceff#1   {\ensuremath{\mathcal{C}_{#1}^{\mathrm{(eff)}}}\xspace}        
\def\Cpeff#1  {\ensuremath{\mathcal{C}_{#1}^{'\mathrm{(eff)}}}\xspace}       
\def\Ope#1    {\ensuremath{\mathcal{O}_{#1}}\xspace}                       
\def\Opep#1   {\ensuremath{\mathcal{O}_{#1}^{'}}\xspace}                    

\newcommand{\nospaceunit}[1]{\ensuremath{\text{#1}}}       
\newcommand{\aunit}[1]{\ensuremath{\text{\,#1}}}       

\newcommand{\tev}{\aunit{Te\kern -0.1em V}\xspace}
\newcommand{\gev}{\aunit{Ge\kern -0.1em V}\xspace}
\newcommand{\mev}{\aunit{Me\kern -0.1em V}\xspace}
\newcommand{\kev}{\aunit{ke\kern -0.1em V}\xspace}
\newcommand{\ev}{\aunit{e\kern -0.1em V}\xspace}
 
\newcommand{\mevc}{\ensuremath{\aunit{Me\kern -0.1em V\!/}c}\xspace}
\newcommand{\gevc}{\ensuremath{\aunit{Ge\kern -0.1em V\!/}c}\xspace}
\newcommand{\mevcc}{\ensuremath{\aunit{Me\kern -0.1em V\!/}c^2}\xspace}
\newcommand{\gevcc}{\ensuremath{\aunit{Ge\kern -0.1em V\!/}c^2}\xspace}

\def\mum  {\ensuremath{\,\upmu\nospaceunit{m}}\xspace}

\def\fb   {\ensuremath{\aunit{fb}}\xspace}
\def\invfb   {\ensuremath{\fb^{-1}}\xspace}

\newcommand{\chisq}{\ensuremath{\chi^2}\xspace}

\newcommand{\chisqip}{\ensuremath{\chi^2_{\text{IP}}}\xspace}

\def\gsim{{~\raise.15em\hbox{$>$}\kern-.85em
          \lower.35em\hbox{$\sim$}~}\xspace}
\def\lsim{{~\raise.15em\hbox{$<$}\kern-.85em
          \lower.35em\hbox{$\sim$}~}\xspace}

\def\sPlot{\mbox{\em sPlot}\xspace}

\def\pt         {\ensuremath{p_{\mathrm{T}}}\xspace}

\def\ptot       {\ensuremath{p}\xspace}

\def\evtgen     {\mbox{\textsc{EvtGen}}\xspace}

\def\geant      {\mbox{\textsc{Geant4}}\xspace}

\def\photos     {\mbox{\textsc{Photos}}\xspace}

\def\pythia     {\mbox{\textsc{Pythia}}\xspace}

\def\tell1  {TELL1\xspace}
\def\ukl1   {UKL1\xspace}

\newcommand{\ie}{\mbox{\itshape i.e.}\xspace}

\newcommand{\lhcborcid}[1]{\href{https://orcid.org/#1}{\hspace*{0.1em}\raisebox{-0.45ex}{\includegraphics[width=1em]{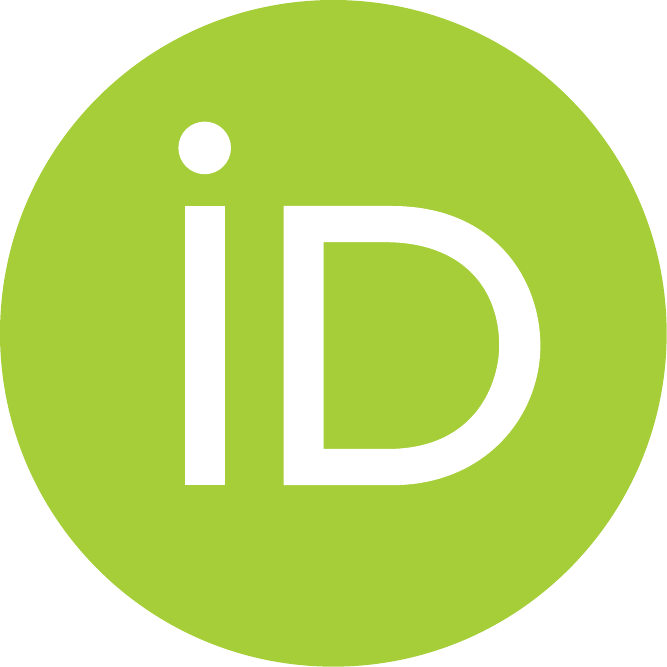}}}}

\usepackage{cite} 
\usepackage{mciteplus}

\newcommand{\mygevc}{\ensuremath{{\mathrm{Ge\kern -0.1em V\!/}c}}\xspace}
\newcommand{\xx}{\ensuremath{\kern 0.5em }}

\def\kpi{\ensuremath{\Kp \pim}\xspace}
\def\pik{\ensuremath{\pip \Km}\xspace}
\def\ppbar{\ensuremath{\proton \antiproton}\xspace}

\newcommand{\BDpp}{\ensuremath{\Bd \to \proton \antiproton}\xspace}
\newcommand{\BSpp}{\ensuremath{\Bs \to \proton \antiproton}\xspace}
\newcommand{\BBspp}{\ensuremath{\Bds \to \proton \antiproton}\xspace}
\newcommand{\BDkpi}{\ensuremath{\Bd \to \Kp \pim}\xspace}

\newcommand{\BSpik}{\ensuremath{\Bs \to \pip \Km}\xspace}

\newcommand{\LBpk}{\ensuremath{\Lb \to p\Km}\xspace}
\newcommand{\LBppi}{\ensuremath{\Lb \to p\pim}\xspace}

\newcommand{\BDpipi}{\ensuremath{\Bd \to \pip \pim}\xspace}
\newcommand{\BSkk}{\ensuremath{\Bs \to \Kp \Km}\xspace}

\newcommand{\BUppk}{\ensuremath{\Bp \to \proton \antiproton \Kp}\xspace}

\newcommand{\BUplb}{\ensuremath{\Bu \to p\Lbar}\xspace}

\def\pp {\ensuremath{pp}\xspace}

\def\sPlot{\mbox{\em sPlot}\xspace}

\usepackage{multirow} 
\usepackage{booktabs} 
\usepackage{rotating}

\newcommand{\BUpLam}{\ensuremath{\Bu \to p\Lbar(1520)}\xspace} 

\usepackage{longtable} 

\begin{document}

\renewcommand{\thefootnote}{\fnsymbol{footnote}}
\setcounter{footnote}{1}


\begin{titlepage}
\pagenumbering{roman}

\vspace*{-1.5cm}
\centerline{\large EUROPEAN ORGANIZATION FOR NUCLEAR RESEARCH (CERN)}
\vspace*{1.5cm}
\noindent
\begin{tabular*}{\linewidth}{lc@{\extracolsep{\fill}}r@{\extracolsep{0pt}}}
\ifthenelse{\boolean{pdflatex}}
{\vspace*{-1.5cm}\mbox{\!\!\!\includegraphics[width=.14\textwidth]{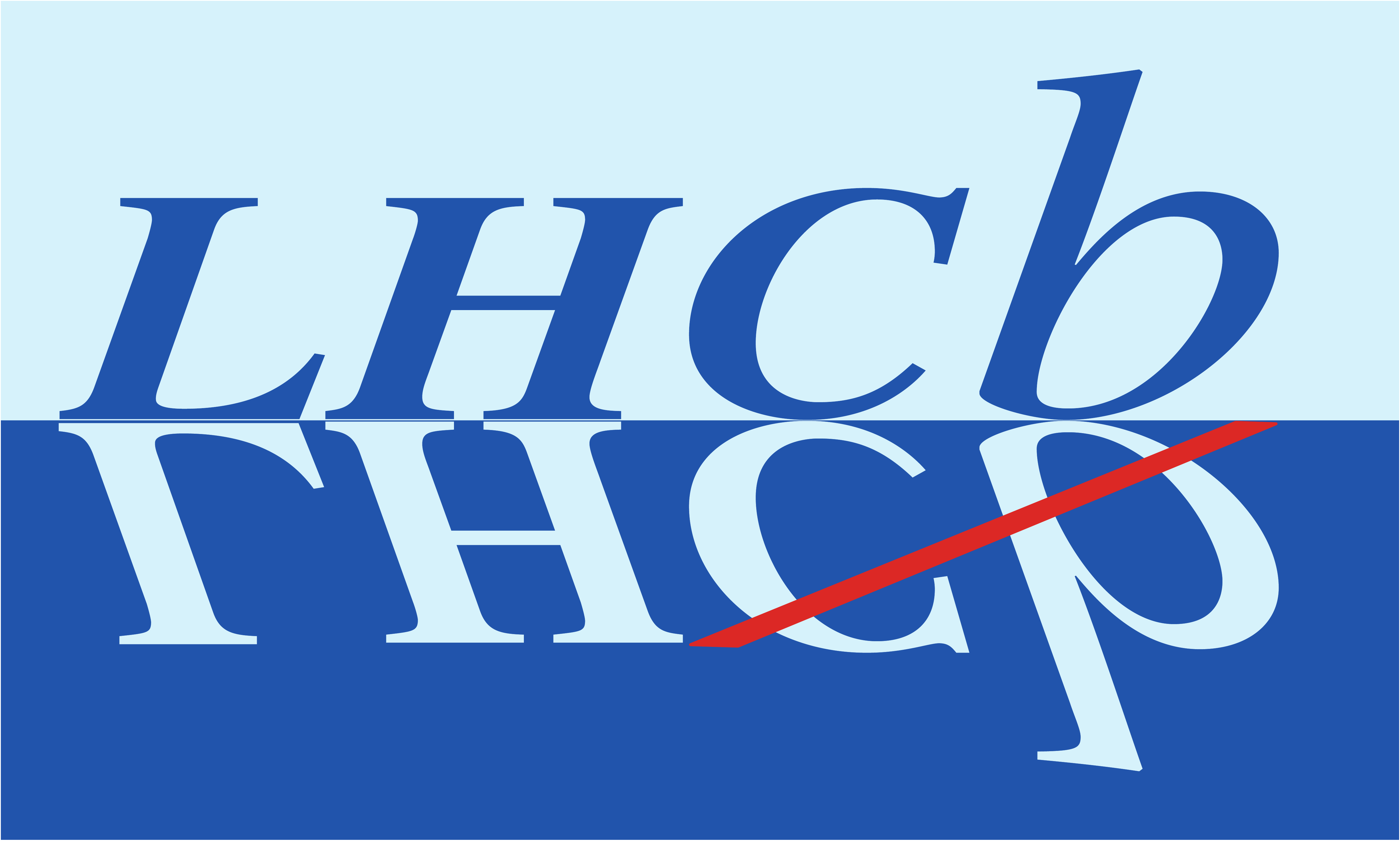}} & &}%
{\vspace*{-1.2cm}\mbox{\!\!\!\includegraphics[width=.12\textwidth]{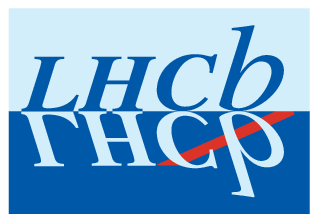}} & &}%
\\
 & & CERN-EP-2022-102 \\  
 & & LHCb-PAPER-2022-004 \\  
 & & \today \\ 
 & & \\
\end{tabular*}

\vspace*{2.5cm}

{\normalfont\bfseries\boldmath\huge
\begin{center}
  \papertitle 
\end{center}
}

\vspace*{2.0cm}

\begin{center}
\paperauthors\footnote{Authors are listed at the end of this paper.}
\end{center}

\vspace{\fill}

\begin{abstract}
  \noindent
A search for the rare hadronic decay \BSpp is performed using proton-proton collision data recorded by the \lhcb experiment at a center-of-mass energy of 13\tev,
corresponding to an integrated luminosity of 6\invfb.
No evidence of the decay is found and an upper limit on its branching fraction is set at $\BR(\BSpp) < 4.4~(5.1) \times 10^{-9}$ at 90\%~(95\%) confidence level;
this is currently the world's best upper limit.
The decay mode \BDpp is measured with very large significance, confirming the first observation by the \lhcb experiment in 2017.
The branching fraction is determined to be
$\BR(\BDpp) = \rm (1.27 \pm 0.15 \pm 0.05 \pm 0.04) \times 10^{-8}$,
where the first uncertainty is statistical, the second is systematic and
the third is due to the external branching fraction of the normalization channel \BDkpi.
The combination of the two \lhcb measurements of the \BDpp branching fraction yields
$\BR(\BDpp) = \rm (1.27 \pm 0.13 \pm 0.05 \pm 0.03) \times 10^{-8}$.
\end{abstract}

\vspace*{2.0cm}

\begin{center}
   Published in Phys.~Rev.~D108~(2023)~012007
\end{center}

\vspace{\fill}

{\footnotesize 
\centerline{\copyright~\papercopyright. \href{\paperlicenceurl}{\paperlicence}.}}
\vspace*{2mm}

\end{titlepage}


\newpage
\setcounter{page}{2}
\mbox{~}


\renewcommand{\thefootnote}{\arabic{footnote}}
\setcounter{footnote}{0}

\cleardoublepage


\pagestyle{plain} 
\setcounter{page}{1}
\pagenumbering{arabic}

\section{Introduction}
\label{sec:introduction}
Exclusive decays of \B mesons to final states containing baryons as well as mesons have been studied since the late 1990s~\cite{Fu:1996qt}.
Differences in the way \B mesons decay to baryonic versus purely mesonic final states have been found since the first experimental measurements became available:
multi-body baryonic decay modes (such as \BUppk) can have their branching fractions significantly increased owing to a threshold enhancement effect in the baryon-antibaryon invariant mass spectrum~\cite{Hou:2000bz, Bevan:2014iga, Hsiao:2019wyd,Huang:2021-review}.
On the other hand, two-body baryonic decays (such as \BBspp) are suppressed~\cite{Hsiao:2014zza, Cheng:2014qxa, Huang:2021-review}. 
To date, only three charmless two-body baryonic decays have been observed, namely the
\BUpLam~\cite{LHCb-PAPER-2013-031}, 
\BUplb~\cite{LHCb-PAPER-2016-048} and 
\BDpp~\cite{LHCb-PAPER-2017-022} modes.

In the Standard Model (SM), the dominant contributions to the \BDpp decay amplitude involve tree-level \bquark \to~\uquark transitions through factorizable \W-exchange and non-factorizable internal \W-emission. 
Other electroweak contributions such as penguin annihilation have a small influence on the decay amplitude.
References~\cite{w-exchange-1,Bevan:2014iga,w-exchange-3,w-exchange-4} indicate that the contribution from \W-exchange or annihilation can be neglected due to helicity suppression.
There is no non-factorizable contribution to the \BSpp decay. Therefore, Refs.~\cite{w-exchange-4,Chua:2016aqy} point out that the penguin-level gluon-exchange and annihilation contributions could be neglected, thus leading to the branching fraction $\BR(\BSpp)$ to be vanishingly small.
Reference~\cite{Hsiao:2019wyd} instead argues that contributions from exchange or annihilation processes could yield $\BR(\BSpp) \sim 10^{-8}$, which is experimentally accessible. 
Consequently, a study of the suppressed \BSpp decay provides valuable information on the role of exchange and annihilation diagrams in baryonic \B decays.

This paper presents a search for the rare baryonic decay mode \BSpp using proton-proton ($pp$) collision data from years
2015--2018 collected by the \lhcb experiment at a center-of-mass energy of 13\tev, corresponding to an integrated luminosity of 6\invfb. 
Prior to this, the \aleph, \cleo, \delphi, \belle, \babar and \lhcb collaborations searched for the \BBspp decays~\cite{LHCb-PAPER-2017-022,B2ppbar-1-ALEPH:1996obw,B2ppbar-1-DELPHI:1995ngp,B2ppbar-3-CLEO:2003oxc,B2ppbar-3-Belle:2007gob,B2ppbar-3-BaBar:2004gwd}.
In 2017 the \lhcb experiment reported the first observation of the \BDpp decay and a measurement of its branching fraction of $\BR(\BDpp) = (1.25 \pm 0.32) \times 10^{-8}$~\cite{LHCb-PAPER-2017-022}. That analysis is based on the dataset recorded at center-of-mass energies of 7 and 8\tev, corresponding to an integrated luminosity of 3\invfb.
The most stringent upper limit of $\BR(\BSpp) < 1.5 \times 10^{-8}$ at 90\% confidence level was set on the branching fraction of the \BSpp decay~\cite{LHCb-PAPER-2017-022}.

In this analysis, the \BDpp and \BSpp yields are determined from a fit to the data and normalized to the yield of \BDkpi decays, which have a similar topology and a precisely measured branching fraction. The use of a normalization channel reduces systematic uncertainties associated to reconstruction and selection efficiencies when determining branching fractions.
The branching fractions of the signal decays are measured using the formula
\begin{equation}
\label{eq:br}
\BR(\BBspp) = \frac{N(\BBspp)}{N(\BDkpi)} \times
\frac{\varepsilon_{\BDkpi}}{\varepsilon_{\BBspp}} \times \BR(\BDkpi) \times \frac{f_d}{f_{d(s)}}\,,
\end{equation}
where $f_s/f_d = 0.2539 \pm 0.0079$ (included only for the \Bs mode) is the ratio of probabilities for \bquark-quark hadronization into \Bd and \Bs mesons~\cite{LHCb-PAPER-2020-046},
$N$ represents the measured yields,
and $\varepsilon$ stands for the detection efficiency including the geometrical acceptance of the \lhcb detector.
The notation \BBspp denotes either \BDpp or \BSpp. 
The inclusion of charged-conjugate processes is implied, unless otherwise stated.

\section{The \lhcb detector and data samples}
\label{sec:DetectorAndDataset}
The \lhcb detector~\cite{LHCb-DP-2008-001,LHCb-DP-2014-002} is a single-arm forward spectrometer covering the pseudorapidity range $2<\eta <5$, designed for the study of particles containing \bquark or \cquark quarks.
The detector includes a high-precision tracking system consisting of a silicon-strip vertex detector surrounding the \pp interaction region~\cite{LHCb-DP-2014-001}, a large-area silicon-strip detector located upstream of a dipole magnet with a bending power of about $4{\mathrm{\,Tm}}$, and three stations of silicon-strip detectors and straw drift tubes~\cite{LHCb-DP-2017-001} placed downstream of the magnet.
The tracking system provides a measurement of the momentum, \ptot, of charged particles with relative uncertainty that varies from 0.5\% at low momentum to 1.0\% at 200\gevc.
The minimum distance of a track to a primary $pp$ collision vertex (PV), the impact parameter (IP), is measured with resolution of $(15+29/\pt)\mum$, where \pt is the component of the momentum transverse to the beam, in~\gevc.
Different species of charged hadrons are distinguished using information from two ring-imaging Cherenkov detectors~\cite{LHCb-DP-2012-003}.
Photons, electrons and hadrons are identified by a calorimeter system consisting of scintillating-pad and preshower detectors, an electromagnetic and a hadronic calorimeter~\cite{LHCb-DP-2020-001}.
The resolution of the energy measurement in the electromagnetic calorimeter varies from around 12\% at 3\gev to around 5\% at 80\gev.
Muons are identified by a system composed of alternating layers of iron and multiwire proportional chambers~\cite{LHCb-DP-2012-002}.

The online event selection is performed by a trigger~\cite{LHCb-DP-2012-004}, which consists of a hardware stage, based on information from the calorimeter and muon systems, followed by a software stage, which applies a full event reconstruction.
At the hardware trigger stage, candidates are required to have a muon with high \pt or a hadron, photon or electron with high transverse energy in the calorimeters.
For hadrons, the transverse energy threshold is 3.5\gev.
The software trigger requires a two-track secondary vertex with significant displacement from any primary \pp interaction vertex.
At least one charged particle must have significant transverse momentum and be inconsistent with originating from a PV.
A multivariate algorithm based on the boosted decision tree technique is used for the identification of secondary vertices consistent with the decay of a \bquark hadron~\cite{BBDT}.

Simulated samples are used to study the properties of the signal, normalization and background channels. 
Proton-proton collisions are generated by \pythia~\cite{Sjostrand:2007gs} with a specific \lhcb configuration~\cite{LHCb-PROC-2010-056}.
Decays of unstable particles are described by \evtgen~\cite{Lange:2001uf}, in which final-state radiation is generated using \photos~\cite{Photos-2}.
The interactions of the generated particles with the detector material, and their responses, are implemented using the \geant toolkit~\cite{Agostinelli:2002hh,Allison:2006ve}, as described in Ref.~\cite{LHCb-PROC-2011-006}.

\section{Candidate selection}
\label{sec:CandidateSelection}
Candidates for both the \BBspp signal channels and the normalization mode \mbox{\BDkpi} are required to pass a set of initial selection criteria, which include requirements on particle identification (PID), and are then further selected using a multivariate classifier relying on a boosted decision tree (BDT) algorithm~\cite{BDT-algorithm}.
To avoid experimenters' bias, the $m(\ppbar)$ invariant mass range [5230, 5417]~\mevcc of the \ppbar signal candidates was not inspected until all the analysis procedures were finalized.
This range corresponds to the combined range spanning approximately three times the invariant mass resolution ($\pm 50$\mevcc) around the known \Bd and \Bs masses~\cite{PDG2020}. 

In the initial selection the \Bds decay products are associated to high-quality tracks with large \chisqip with respect to any PV, where the \chisqip is defined as the difference between the vertex-fit \chisq of a PV reconstructed with and without the track in question.
The \pt of the decay products is required to be between 900\mevc and 40\gevc and at least one of the decay products is required to have \pt $> 2100$\mevc.
Furthermore, the decay products are required to satisfy $\ptot <500\gevc$ and $1.5 < \eta < 5.0$.
Both particles are required to pass a loose PID selection based primarily on information from the Cherenkov detectors, with fiducial cuts in momentum and pseudorapidity.
The angle $\theta_{B}$ between the momentum vector of the \Bds candidate and the line connecting the associated PV and the decay vertex of the candidate must be close to zero.
Likewise, the decay vertex of the \Bds candidate is required to be significantly displaced from the PV and its transverse momentum is required to be larger than 500\mevc.
The decay time of the \Bds candidate is limited to be less than ten times the \Bds lifetime.

After the initial selection a tight PID selection is applied to the two final-state particles to remove the contamination from $b$-hadron decays where one or more decay products are misidentified. This PID selection also significantly reduces combinatorial background formed from accidental associations of tracks unrelated to the signal decays.
The optimal PID requirements are found by maximizing the figure of merit $\varepsilon_{\rm sig} / (\alpha/2 + \sqrt{N_{\rm bkg}})$~\cite{Punzi:2003bu}, where $\varepsilon_{\rm sig}$ represents the PID selection efficiency of the signal channel.
The quantity $N_{\rm bkg}$ stands for the expected number of background candidates in the signal region. It is estimated by extrapolating the result of a fit to the invariant mass distribution of the data from the data sideband regions (above and below the signal region, \ie [5080, 5230]~\mevcc and [5417, 5570]~\mevcc).
The term $\alpha$ specifies the target signal significance in units of standard deviations; for this analysis a value $\alpha = 3$ is chosen.

Following the PID selection a multivariate classifier, based on a BDT algorithm and implemented with the TMVA package~\cite{Hocker:2007ht,TMVA4}, is applied as the final stage of the entire selection chain.
The main purpose is to keep as much of the \BBspp signals as possible whilst removing combinatorial background.
Ten input quantities for the BDT classifier are chosen to maximize signal-to-background discrimination: 
two isolation variables~\cite{Aaltonen:2013as, Leo:2012us} used to reject background candidates from decays with additional charged or neutral particles in the final state;
the \chisqip of the \Bds candidate with respect to the PV; the \Bds candidate vertex fit \chisq;
the cosine of the direction angle $\theta_{B}$ of the \Bds candidate;
the separation between the two outgoing proton tracks at their point of closest approach;
the smallest of the \chisqip with respect to the PV of the candidate final-state particles;
the sum of transverse momenta of the candidate final-state particles;
the smallest of the pseudorapidities of the candidate final-state particles;
the cosine of the angle between the proton momentum in the \Bds rest-frame and the vector perpendicular to the \Bds momentum and the beam axis.
The BDT classifier is trained using fully selected \BDpp simulated events as signal and sideband data as background.
The sideband data are taken from across the wider \ppbar invariant mass spectrum ranging from 5080 to 5570\mevcc, excluding candidates within the signal region [5230, 5417]\mevcc.
The requirement on the BDT response is optimized using the same figure of merit as that used in the optimization of the PID selection.
The independence of the BDT response on the two-body invariant mass is verified.
The same BDT selection is applied to the normalization decay candidates.
In this way, systematic effects on the ratio of BDT selection efficiencies entering the relative branching fraction calculation cancel to a large extent.

Possible sources of background to the \ppbar and \kpi spectra are studied with simulated samples.
These background components contain partially reconstructed decays with one or more particles from the \bquark-hadron decays escaping detection, and two-body \bquark-hadron decays where one or both final-state particles are misidentified.
The relevant backgrounds in both mass spectra are detailed in the following section.

\section{Signal yield determination}
\label{sec:sigYieldDetermination}
The yields of the signal and background candidates after full selection in the signal and control model spectra are determined with unbinned maximum likelihood fits to the \ppbar and \kpi invariant-mass distributions, respectively.

\subsection{Mass fit to the control channel \boldmath{\BDkpi}}
\begin{figure}[tbp]
    \centering
    \includegraphics[width=0.98\textwidth]{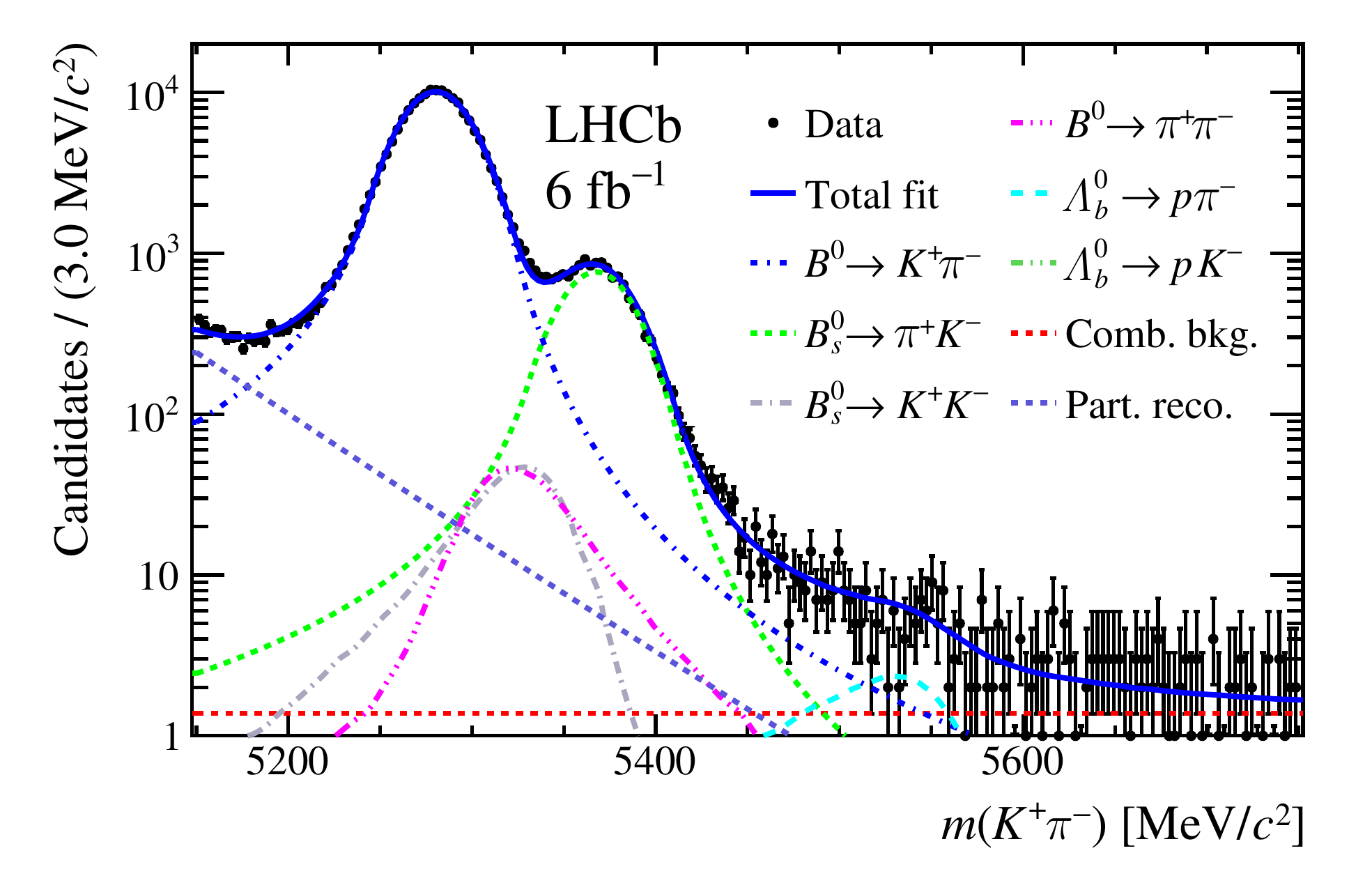}
    \caption{Invariant-mass distribution of \BDkpi candidates. 
    The fit result is shown together with each fit model component.}    
    \label{fig:massFitForKpi}
\end{figure}

The distribution of the invariant mass $m(\Kp\pim)$ of selected \BDkpi candidates, shown in Figure~\ref{fig:massFitForKpi},
is described with the following components: 
the \BDkpi control channel and \BSpik; 
the background contributions due to the decays \mbox{\BSkk}, \BDpipi, \LBppi, and \LBpk when at least one of the final-state particles is misidentified;
partially reconstructed background from \bquark-hadron decays, such as \mbox{$\Bu \to \proton \antiproton \Kp$};
and the combinatorial background. 
Contamination from other decays is treated as a systematic uncertainty.
The fits are performed over the \kpi invariant-mass range from 5150 to 5750\mevcc.
The lower mass limit is chosen to reject the majority of the partially reconstructed background,
thus including only the high-mass tail of the contribution, which comes into the \BDkpi signal region.
The upper mass limit is chosen to provide adequate constraints on the model parameters of the combinatorial background model.

The \BTohh signal mass distributions are modeled with a double-sided Crystal Ball (DSCB) function~\cite{Skwarnicki:1986xj},
characterized by exponential tails on both sides of a Gaussian core.
The mean and the width of the Gaussian core function describing the \BDkpi signal channel are free to vary in the fit. 
The difference between the peak positions of the \BDkpi and \BSpik signal decays is constrained to the known value of 87.35\mevcc~\cite{PDG2020}.
The core width of the \BSpik decay mode is related to that of \BDkpi by a scaling factor of 1.01 as determined from the simulation.
The four DSCB tail parameters are also extracted from the simulation.

Misidentified \BTohh and $\Lb \to ph^-$ decays (where $h^{(')}$ denotes a \pion or a \kaon meson) with expected yields greater than 0.1\% of the \BDkpi yield are included individually in the fit.
The shapes are modeled with non-parametric distributions, extracted from simulated samples, reconstructed and selected under the \kpi mass hypothesis, and smoothed using a kernel method~\cite{Cranmer:2000du} to suppress statistical noise due to finite simulated samples.
The relative fractions of these background sources depend on the branching fractions, $b$-hadron hadronization probabilities, and misidentification rates. 
The fractions are left to vary in the fit, but their values are constrained with Gaussian functions: the means are set to the products of the three described factors and the widths defined to be equal to the combined uncertainties on the three factors.

The shape of the partially reconstructed background component in the \kpi mass hypothesis, which represents several background decay modes, is determined from a mixture of simulated samples.
Each mode is assigned a weight given by its relative branching fraction, the hadronization fraction and selection efficiency.
The spectrum of the ensemble of partially reconstructed background components is well modeled with the sum of two exponential functions.
The overall yield is extracted from the fit to the data, while all other parameters are fixed from simulation.
The invariant-mass distribution of the combinatorial background is described with an exponential function of which the slope parameter is determined in the fit.

The number of signal candidates in the normalization channel is extracted from unbinned maximum likelihood fits to the \kpi and \pik invariant-mass spectra, shown together in Figure~\ref{fig:massFitForKpi}, yielding $N_{\BDkpi} = 179902 \pm 450$ signal decays, where the uncertainty is statistical only. Both fits yield compatible results for the signal shape parameters and the fractions of misidentified backgrounds.
The difference in yields from the individual fits corresponds to a raw asymmetry that is found to be consistent with the result in a previous \lhcb analysis~\cite{LHCb-PAPER-2013-018}.

\subsection{Mass fit to \boldmath{\BBspp}}
\begin{figure}[tbp]
    \centering
    \includegraphics[width=0.95\textwidth]{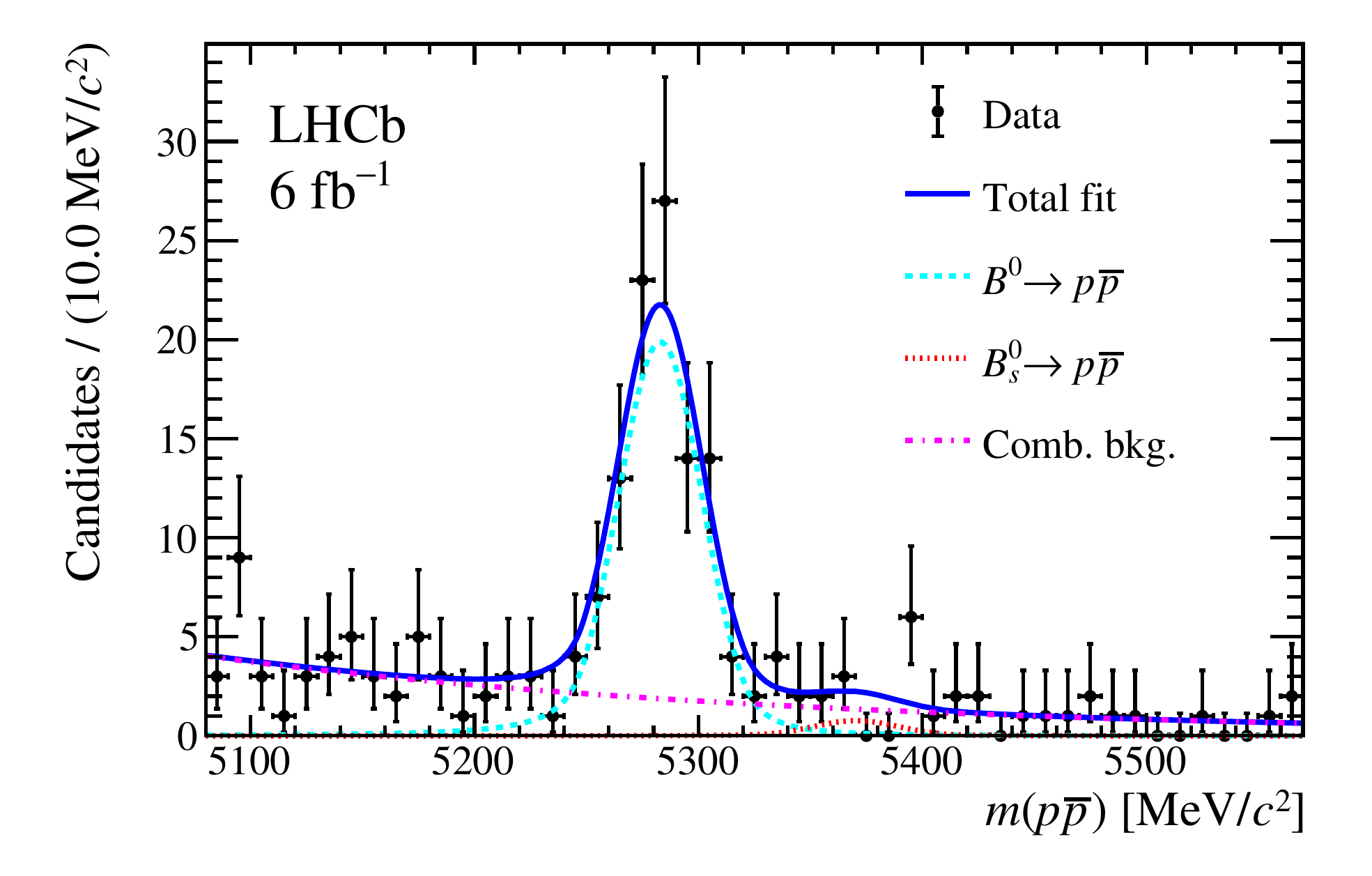}
    \caption{Invariant-mass distribution of \ppbar candidates with each fit model component.}
    \label{fig:ppMassFit}
\end{figure}

The \ppbar invariant-mass distribution for selected candidates is shown in Figure~\ref{fig:ppMassFit}. The shape of this distribution is modeled with three components, namely the \BDpp and \BSpp signals, and  combinatorial background.
Contributions from specific physics background sources are determined to be negligible after the full event selection has been applied.
The \BBspp signal shapes are modeled with DSCB functions, as the \BDkpi normalization channel.
The core widths are taken from simulation with scaling factors to account for differences in the resolution between data and simulated samples as determined from \BDkpi candidates.
The peak of the \BDpp DSCB is free to vary in the fit whereas the peak of the \BSpp DSCB is again constrained by the known \Bs--\Bd mass difference~\cite{PDG2020}.
The tail parameters of both \BBspp modes are fixed from simulation.
The mass distribution of the combinatorial background is described by an exponential function with the gradient as the only free parameter in the fit. 

The yields of the \BBspp decays are $N(\BDpp) = 98 \pm 11$ and $N(\BSpp) = 4 \pm 5$ candidates, where the uncertainties are statistical.
The significance of each of the signals is calculated from the change in the logarithm of the likelihood between fits with and without the signal component, following Wilks’ theorem~\cite{Wilks:1938dza}.
The statistical signal significance is $16.2\,\sigma$ and $0.9\,\sigma$ for the \BDpp and \BSpp decays, respectively.
This measurement hence confirms the observation of the \BDpp decay previously published by the \lhcb collaboration~\cite{LHCb-PAPER-2017-022}, while no evidence for the \BSpp decay is found.

\section{Systematic uncertainties}
\label{sec:Syst}
Systematic uncertainties cancel to a large extent because the branching fractions are measured relative to that of a topologically identical normalization mode.
The main sources of uncertainty on the relative branching fraction measurement are the selection efficiencies, the \BDkpi branching fraction, $f_s/f_d$, and the fit model. They are summarized in Table~\ref{tab:syst_from_tot} (except for the contributions from the normalization mode) and hereafter discussed.

\begin{table}[tbp]
    \centering
    \caption{Systematic uncertainties on the \BBspp branching fractions, in percentage (\%), coming from the \BDpp and \BSpp contributions in Equation~\ref{eq:br}. The totals correspond to the quadratic sums of each column.}
    \vspace{0.2cm}
    \label{tab:syst_from_tot}
    \begin{tabular}{l|c|c} \hline
        Source of systematic uncertainties &  \BDpp &  \BSpp \\ \hline
        $f_s/f_d$ & - & 3.1 \\  
        L0 trigger efficiency & 1.0 &  1.0 \\
        Selection efficiency relative to \BDkpi & 2.0 & 2.0 \\
        Tracking efficiency & 1.9 &  1.9 \\
        PID efficiency & 2.4 & 2.4 \\
        Fit model & 1.0 & 22.0 \\ \hline
        Total & 3.9 & 22.5 \\ \hline
    \end{tabular}
\end{table}

The branching fraction of the normalization mode, $\BR(\BDkpi) = (1.96 \pm 0.05) \times 10^{-5}$, is taken from Ref.~\cite{PDG2020}.
Its relative uncertainty of 2.6\% is taken as a systematic uncertainty on the \BBspp branching fractions.
For the \BSpp branching fraction measurement an extra 3.1\% relative uncertainty comes from the uncertainty on the ratio of fragmentation fractions $f_s/f_d$.

The trigger efficiencies are assessed with simulation.
The \pt distributions of the decay final-state particles are well described in simulation and all final states comprise the same number of tracks.
Neglecting very small $p$ and \pt differences between the \BDpp and \BSpp modes, the ratios of $\BDkpi/\BDpp$ and $\BDkpi/\BSpp$ trigger efficiencies are identical within uncertainties.
No uncertainties are considered for the software triggers given that the dominant source of systematic uncertainty is the reconstruction of charged particles, taken into account below as tracking systematic uncertainties.
The main contribution to the uncertainties on the trigger efficiencies comes from the imperfect description of the hardware trigger in the simulation.
The accuracy of the efficiency determinations is assessed exploiting trigger calibration data subdivided by particle charge and momentum distribution
to obtain efficiency maps for protons, kaons and pions individually.
The effect is studied with simulated toys for the signals and the normalization mode, accounting for observed variations in the efficiency maps.
A systematic uncertainty on the relative \lone efficiencies of 1.0\% is assigned for both signal modes based on the differences observed with respect to the default determination of the efficiencies.

The \BDkpi decay mode is used to assess a systematic uncertainty related to the selection.
The discrepancies between data and simulation are assessed with \BDkpi signal distributions obtained from weighted collision data, using the \sPlot technique with $m(\Kp\pim)$ as discriminating variable~\cite{Pivk:2004ty}.
The 2.0\% level of disagreement between simulation and weighted data is assigned as the systematic uncertainty for the ratios of the \BBspp to \BDkpi selection efficiencies.

Differences in the reconstruction of the decay particles may arise to some extent given that the final states of the signal modes contain only baryons whereas the final state of the normalization mode contains only mesons.
The track reconstruction efficiencies of the \BBspp and \BDkpi decays are corrected with tracking calibration samples~\cite{LHCb-DP-2013-002}.
A systematic uncertainty of 1.9\% is assigned on the relative efficiency for reconstructing \BBspp decays versus \BDkpi decays.

The systematic uncertainties on the PID efficiencies are determined using a method based on resampling,
alternative to the default method of correcting on a particle-by-particle basis the PID variables in simulation
in order to match the PID performance in data.
The data-driven PID resampling method effectively reweights the PID distributions in simulation exploiting calibration data samples,
from which the PID selection efficiencies are evaluated.
The largest discrepancy observed between both methods, which is 2.4\%, is assigned as the overall systematic uncertainty due to the relative PID selection efficiencies for both \BDpp and \BSpp.

The fit systematic uncertainties originate from the finite accuracy of the mass models
and from the uncertainties on the fixed parameters in the fits.
They are investigated with pseudoexperiments in which the fixed parameters of the models are varied within their uncertainties or the combinatorial background is modeled with an alternative (second-order) polynomial function. The observed signal yields are used as the input means for the Poisson distributions.
The fit systematic uncertainties for the \BDpp and \BSpp signals are 1.0\% and 22.0\%, respectively; for the normalization mode they amount to 1.0\%.
The overall systematic uncertainties on the \BDpp, \BSpp branching fractions coming from the \BDpp, \BSpp and \BDkpi contributions in Equation~\ref{eq:br} are 3.9\%, 22.5\% and 2.8\%, respectively, combining all sources in quadrature.

\section{Results and conclusions}
\label{sec:Results}
\begin{figure}[tbh]
    \centering
    \includegraphics[width=0.9\textwidth]{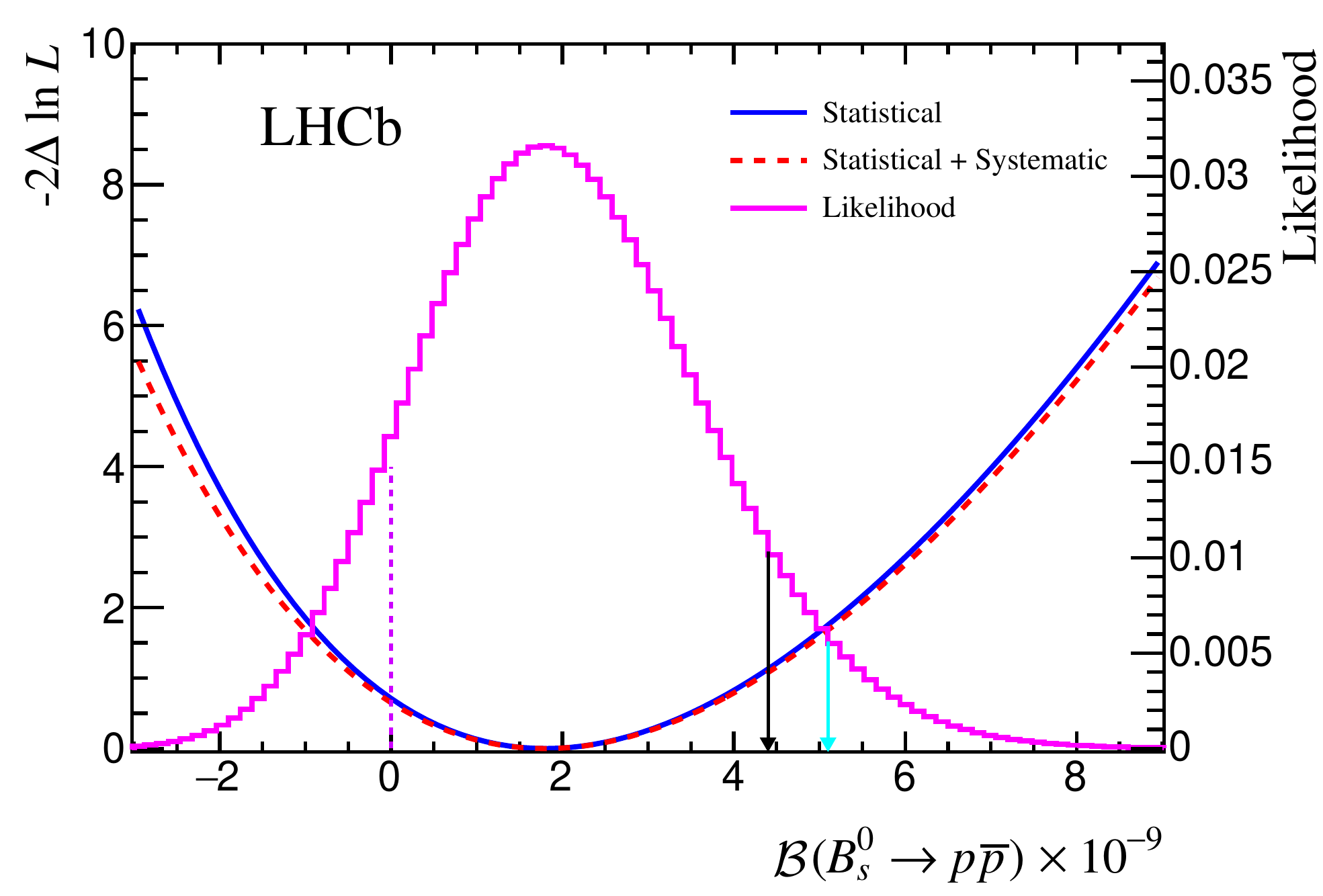}
    \caption{Negative logarithm of the profile likelihoods (${\rm -2 \Delta ln}L$) and the corresponding likelihood as a function of the \BSpp branching fraction. 
    The solid blue line represents the statistical-only profile, whereas the dashed red line includes the systematic uncertainties.
    The solid magenta line shows the distribution of likelihood.
    The dashed violet vertical line at 0 shows the start point of the integral.  
    The black (cyan) arrow shows the upper limits at 90\%  (95\%) confidence level. }
    \label{fig:upperLimitBs}
\end{figure}

In summary, an updated study of the rare two-body charmless baryonic decays \BBspp is performed using the full proton-proton collision data sample collected with the \lhcb detector, during 2015--2018, at a center-of-mass energy of 13\tev, corresponding to an integrated luminosity of approximately 6\invfb.
No evidence for a \BSpp signal is found and an upper limit (UL) at 90\% (95\%) confidence level (CL) is computed by integrating the likelihood as a function of the branching fraction,
where the likelihood probability distribution function 
is obtained from the fit profile likelihood 
using a flat prior on the branching fraction: 
\begin{equation*}
\int_{0}^{\rm UL} L({\cal B}|\theta) \textup{d}{\cal B} = \textup{(CL)}\int_{0}^{\infty} L({\cal B}|\theta) \textup{d}{\cal B}\,.
\end{equation*}
Figure~\ref{fig:upperLimitBs} shows the resulting likelihood as a function of the branching fraction with combined statistical and systematic uncertainties, 
where the effect of the systematic uncertainties is accounted for convolving the statistical-only likelihood with a Gaussian function of mean zero and width equal to the fit systematic uncertainty.
The 90\% (95\%) upper limit on the \BSpp decay branching fraction is set at
\begin{equation*}
\BR(\BSpp) < 4.4~(5.1) \times 10^{-9} \rm~at~90\%~(95\%)~CL.
\end{equation*}
This world's best upper limit supersedes and improves the previous limit by approximately a factor of four, and constrains contributions from beyond SM processes.

The \BDpp decay mode observation is confirmed with very large significance.
Using the measured quantities of Equation~\ref{eq:br}
the branching fraction of the \BDpp decay is measured to be
\begin{equation*}
\BR(\BDpp) = \rm (1.27 \pm 0.15 \pm 0.05 \pm 0.04) \times 10^{-8}\,,
\end{equation*}
where the first uncertainty is statistical, the second is systematic and 
the third is due to the uncertainty on the branching fraction of the normalization channel.

The combination of this measurement with the previous \lhcb measurement~\cite{LHCb-PAPER-2017-022} (data sample from years 2011--2012) of the \BDpp branching fraction is based on the minimization of a $\chi^2$ dependent on the two measurements and their uncertainties.
It provides
\mbox{$\BR(\BDpp) = \rm (1.27 \pm 0.13 \pm 0.05 \pm 0.03) \times 10^{-8}$},
where correlated sources of systematic uncertainties are assumed as fully correlated and all other uncertainties considered as uncorrelated. The sources of correlated systematic uncertainties are the following: the \BDkpi branching fraction, the relative \lone trigger uncertainties and the mass fit model parameters.

This precise measurement of the \BDpp branching fraction and stringent limit on the \BSpp branching fraction,
together with the experimental knowledge of the \BUplb branching fraction,
improves the extraction of both tree and penguin amplitudes of charmless two-body baryonic $B$ decays~\cite{Chua:2016aqy}.
A possible observation of the very rare decay mode \BSpp and a determination of its branching fraction 
requires future data to be collected by \lhcb.
\section*{Acknowledgements}
%
%
\noindent We express our gratitude to our colleagues in the CERN
accelerator departments for the excellent performance of the LHC. We
thank the technical and administrative staff at the LHCb
institutes.
We acknowledge support from CERN and from the national agencies:
CAPES, CNPq, FAPERJ and FINEP (Brazil); 
MOST and NSFC (China); 
CNRS/IN2P3 (France); 
BMBF, DFG and MPG (Germany); 
INFN (Italy); 
NWO (Netherlands); 
MNiSW and NCN (Poland); 
MEN/IFA (Romania); 
MICINN (Spain); 
SNSF and SER (Switzerland); 
NASU (Ukraine); 
STFC (United Kingdom); 
DOE NP and NSF (USA).
We acknowledge the computing resources that are provided by CERN, IN2P3
(France), KIT and DESY (Germany), INFN (Italy), SURF (Netherlands),
PIC (Spain), GridPP (United Kingdom), 
CSCS (Switzerland), IFIN-HH (Romania), CBPF (Brazil),
Polish WLCG  (Poland) and NERSC (USA).
We are indebted to the communities behind the multiple open-source
software packages on which we depend.
Individual groups or members have received support from
ARC and ARDC (Australia);
Minciencias (Colombia);
AvH Foundation (Germany);
EPLANET, Marie Sk\l{}odowska-Curie Actions and ERC (European Union);
A*MIDEX, ANR, IPhU and Labex P2IO, and R\'{e}gion Auvergne-Rh\^{o}ne-Alpes (France);
Key Research Program of Frontier Sciences of CAS, CAS PIFI, CAS CCEPP, 
Fundamental Research Funds for the Central Universities, 
and Sci. \& Tech. Program of Guangzhou (China);
GVA, XuntaGal, GENCAT and Prog.~Atracci\'on Talento, CM (Spain);
SRC (Sweden);
the Leverhulme Trust, the Royal Society
 and UKRI (United Kingdom).

\clearpage
\addcontentsline{toc}{section}{References}
\setboolean{inbibliography}{true}
\bibliographystyle{LHCb}
\bibliography{main,LHCb-PAPER,LHCb-CONF,LHCb-DP,LHCb-TDR,standard,MyBib}

\newpage
\centerline
{\large\bf LHCb collaboration}
\begin
{flushleft}
\small
R.~Aaij$^{32}$\lhcborcid{0000-0003-0533-1952},
A.S.W.~Abdelmotteleb$^{50}$\lhcborcid{0000-0001-7905-0542},
C.~Abellan~Beteta$^{44}$,
F.~Abudin{\'e}n$^{50}$\lhcborcid{0000-0002-6737-3528},
T.~Ackernley$^{54}$\lhcborcid{0000-0002-5951-3498},
B.~Adeva$^{40}$\lhcborcid{0000-0001-9756-3712},
M.~Adinolfi$^{48}$\lhcborcid{0000-0002-1326-1264},
H.~Afsharnia$^{9}$,
C.~Agapopoulou$^{13}$\lhcborcid{0000-0002-2368-0147},
C.A.~Aidala$^{76}$\lhcborcid{0000-0001-9540-4988}
% ...

{\footnotesize \it

% ...
$^{9}$Universit{\'e} Clermont Auvergne, CNRS/IN2P3, LPC, Clermont-Ferrand, France\\
% ...
$^{13}$LPNHE, Sorbonne Universit{\'e}, Paris Diderot Sorbonne Paris Cit{\'e}, CNRS/IN2P3, Paris, France\\
% ...
$^{32}$Nikhef National Institute for Subatomic Physics, Amsterdam, Netherlands\\
% ...
$^{40}$Instituto Galego de F{\'\i}sica de Altas Enerx{\'\i}as (IGFAE), Universidade de Santiago de Compostela, Santiago de Compostela, Spain\\
% ...
$^{44}$Physik-Institut, Universit{\"a}t Z{\"u}rich, Z{\"u}rich, Switzerland\\
% ...
$^{48}$H.H. Wills Physics Laboratory, University of Bristol, Bristol, United Kingdom\\
% ...
$^{50}$Department of Physics, University of Warwick, Coventry, United Kingdom\\
% ...
$^{54}$Oliver Lodge Laboratory, University of Liverpool, Liverpool, United Kingdom\\
% ...
$^{62}$Syracuse University, Syracuse, NY, United States\\
% ...
$^{76}$University of Michigan, Ann Arbor, MI, United States, associated to $^{62}$\\
\bigskip
% ...
}
\end{flushleft}

\end{document}